\documentclass[journal]{IEEEtran}

\usepackage{amsmath,graphicx,color,psfrag,mathrsfs,bm}
\usepackage{hyperref}
\usepackage{longtable}          
\usepackage{amsthm}
\usepackage{amssymb}
\usepackage{multicol}
\usepackage{mathtools}
\usepackage{makebox}
\newtheorem{thm}{Theorem}[section]

\newtheorem{assumption}[thm]{Assumption}

\theoremstyle{definition}

\theoremstyle{remark}

\begin{document}
%
\title{Robust Hypothesis Testing with $\alpha$-Divergence}

\author{G\"{o}khan G\"{u}l,~\IEEEmembership{Student~Member,~IEEE,}
        Abdelhak M. Zoubir,~\IEEEmembership{Fellow,~IEEE,}
\thanks{G. G\"{u}l and A. M. Zoubir are with the Signal Processing Group, Institute
of Telecommunications, Technische Universität Darmstadt, 64283, Darmstadt,
Germany (e-mail: ggul@spg.tu-darmstadt.de; zoubir@spg.tu-darmstadt.de)}
\thanks{Manuscript received April 19, 2005; revised January 11, 2007.}}

\maketitle

\begin{abstract}
A robust minimax test for two composite hypotheses, which are determined by the neighborhoods of two nominal distributions with respect to a set of distances - called $\alpha-$divergence distances, is proposed. Sion's minimax theorem is adopted to characterize the saddle value condition. Least favorable distributions, the robust decision rule and the robust likelihood ratio test are derived. If the nominal probability distributions satisfy a symmetry condition, the design procedure is shown to be simplified considerably. The parameters controlling the degree of robustness are bounded from above and the bounds are shown to be resulting from a solution of a set of equations. The simulations performed evaluate and exemplify the theoretical derivations.

\end{abstract}


%
\IEEEpeerreviewmaketitle

\section{Introduction}
\label{sec:intro}
Decision theory has been an active field of research benefiting from contributions from several disciplines, such as economics, engineering, mathematics, or statistics. A decision maker (or a detector) chooses a course of action from several possibilities.
A detector is said to be optimal or to be giving the \it best \rm decision for a particular problem
if the decision rule of interest minimizes (or maximizes) a well defined cost function, e.g., the error probability (or the probability of detection) \cite{levy}.\\
In addition to the fact that decision theory is truly an interdisciplinary subject of research, there are many areas of engineering,
where decision theory finds applications, e.g., radar, sonar, seismology, communications and biomedicine. For some applications, such as image and speech classification or pattern recognition, interest is in a statistical test that performs well on average.
However, for safety oriented applications such as seismology or forest fire detection, as well as for biomedical applications such as early cancer detection from magnetic resonance images or X-ray images, interest is in maximizing the worst case performance because
the consequences of an incorrect decision can be severe \cite{levy}.\\
In general, any practical application of decision theory can be formulated as a hypothesis testing problem.
For binary hypothesis testing, it is assumed that under each hypothesis $\mathcal{H}_i$, the received data $y=(y_1,\ldots,y_n)\in\Omega$ follows a particular distribution $F_i$ corresponding to a density function $f_i$, $i\in\{0,1\}$. A decision rule $\delta$ partitions the whole observation space $\Omega$ into non-overlapping regions corresponding to each hypothesis. The optimality of the decision rule $\delta$ depends on the correctness of the assumption that the data $y$ follows $F_i$. However, in many practical applications either $F_0$ and/or $F_1$ are partially known or are affected by some secondary physical effects that go unmodeled \cite{levy09}.\\
Imprecise knowledge of $F_0$ or $F_1$ leads, in general, to performance degradation and a useful approach is to extend the known model by accepting a set of distributions $\mathcal{F}_i$, under each hypothesis $\mathcal{H}_i$, that are populated by probability distributions $G_i$, which are at the neighborhood of the nominal distribution $F_i$ based on some distance $D$ \cite{levy}.
Under some mild conditions on $D$, it can be shown that the best (error minimizing) decision rule $\hat{\delta}$ for the worst case (error maximizing) pair of probability distributions $(\hat{G_0},\hat{G_1})\in\mathcal{F}_0\times \mathcal{F}_1$ accepts a saddle value. Therefore, such a test design guarantees a certain level of detection at all times. This type of optimization is known as minimax optimization and  the corresponding worst case distributions $(\hat{G_0},\hat{G_1})$ are called least favorable distributions (LFD)s \cite{gul}.\\
The literature in this field is unfortunately not rich. One of the earliest and probably the most crucial work goes back to Huber, who proposed
a robust version of the probability ratio test for the $\epsilon-$contamination and total variation classes of distributions \cite{hube65}. He proved the existence of least favorable distributions and showed that the corresponding robust test was a censored version of the nominal likelihood ratio for both uncertainty classes. In a later work, Huber and Strassen extended the $\epsilon-$contamination neighborhood to a larger class, which includes five different distances as special cases \cite{hube68}. It was also shown that the robust test resulting from this new neighborhood was still a censored likelihood ratio test. Although it was found to be less engineering oriented by Levy \cite{levy}, the largest classes for which similar conclusions have been made was for the $2-$alternating capacities proposed by Huber and Strassen \cite{hube73}.\\
Another approach for robust hypothesis testing was proposed by Dabak and Johnson based on the fact that the choice of measures defining the contamination neighborhoods was arbitrary \cite{dabak}. They chose the relative entropy (KL-divergence) because it is a natural distance between probability measures and therefore a natural way to define the contamination neighborhoods. Somewhat surprisingly, the robust test which minimizes the KL-divergence between the LFDs obtained from the closed balls with respect to the relative entropy distance was not a clipped likelihood ratio test,
but a nominal likelihood ratio test with a modified threshold. It was noted that their approach was not robust for all sample sizes but when Kullback's theorem is valid, that is for a large number of observations \cite{dabak}.
The difference in the robust tests for $\epsilon-$contamination and relative entropy neighborhoods lies in the fact that all the densities in the
class of distributions based on relative entropy are absolutely continuous with respect to the nominal distributions, but not for the case of the $\epsilon-$contamination class.\\
A question left open by Dabak and Johnson was the design of a robust test for a finite number of samples.
Levy answered this question under two assumptions; monotone increasing nominal likelihood ratio and symmetric nominal density functions $(f_0(y)=f_1(-y))$, where $y\in\mathbb{R}$. He implied that a robust test based on the relative entropy would be more suitable for modeling errors rather than outliers, due to the smoothness property (absolute continuity). He also showed that the resulting robust test was neither equivalent to Huber's nor to Dabak's robust test; it was a completely different test \cite{levy09}.\\
Although KL-divergence is a smooth and a natural distance between probability measures, it is not clear why KL divergence should be considered to build uncertainty sets, especially since there are many other divergences, which are also smooth and have nice theoretical properties, e.g. the symmetry property, which KL-divergence does not have. Besides, theoretically nice properties do not always lead to preferable engineering applications, see for example \cite[p.7]{family}. In this respect, KL-divergence can be replaced by the $\alpha-$divergence because $\alpha-$divergence includes uncountably many distances as special cases, e.g. $\chi^2$ distance for $\alpha=2$ \cite{entropy}, it reduces to the KL-divergence as $\alpha\rightarrow 1$ and shares similar theoretical properties with the KL-divergence such as smoothness, convexity or satisfiability of (generalized) Pythagorean inequality \cite{tim}. Moreover, the flexibility provided by the choice of $\alpha$ results in performance improvements in various signal processing applications and implies the sub-optimality of the KL-divergence. For example, in the design of distributed detection networks with power constraints, $\alpha-$divergence is considered as the distance between the probability measures, and error exponents of both kinds are maximized over all $\alpha\in(0,1)$ \cite{tuncel}. In non-negative matrix factorization \cite{cichocki}, and indexing and retrieval \cite{hero}, the optimal value of $\alpha$ (with respect to some objective function) is found to be $1/2$ corresponding to the squared Hellinger distance. In medical applications; e.g. in medical image segmentation \cite{med1}, restoration \cite{med2} and registration \cite{med3}, the $\alpha-$divergence is considered and the optimal value of $\alpha$ is found to be a non-standard value, i.e. a value which does not correspond to any known distance. There are also theoretical works which take advantage of the $\alpha-$divergence in the design of statistical tests. It is reported for parametric models \cite{par1,par2} as well as for non parametric models \cite{par3} that the use of $\alpha-$divergence as the distance between probability measures, again with some non-standard values of $\alpha$, e.g. $\alpha=1.6$ in \cite{par2} and $\alpha=1.3$ or $\alpha=1.5$ in \cite{par3}, leads to promising results. However, non of these aforementioned works have the property of minimax robustness. Furthermore, in non of the aforementioned works, it is possible to adjust the tradeoff between robustness and detection performance. Additionally, the parametric models have a possibly invalid assumption that the actual probability distributions can be represented by a parametric model. This motivates the work in this paper: a minimax robust design of hypothesis testing with the $\alpha-$divergence distance, where the robustness is adjustable with respect to the detection performance by the choice of two robustness parameters, $\epsilon_0$ and $\epsilon_1$.\\
The related literature can be summarized as follows: In \cite{gul}, the symmetry constraint that was imposed in \cite{levy09} was removed, considering the squared Hellinger distance. In \cite{gul3}, the number of non-linear equations that needs to be solved to be able to design the robust test was reduced and a formula from where the maximum robustness parameters could be obtained was derived. In \cite{gul2}, robust approaches were extended
to distributed detection problems where communication from the sensors to the fusion center is constrained. In a recent work \cite{gul5}, based on the KL-divergence, the monotone increasing likelihood ratio constraint was removed.\\
In this paper, A minimax robust test for two composite hypotheses, which are formed by the neighborhoods of two nominal distributions with respect to the $\alpha-$divergence, is designed. It is shown that for any $\alpha$, the corresponding robust test is the same and unique. There is no constraint on the choice of nominal distributions. Therefore, our design generalizes \cite{levy09}. Since the $\alpha-$divergence includes the KL-divergence or the squared Hellinger distance as a special case, cf. \cite{entropy}, our work also generalizes the works in \cite{gul,gul3} and \cite{gul5}. The advantage of considering the $\alpha-$divergence for modeling errors is that it allows the designer to choose a single parameter that accounts for the distance without carrying out tedious steps of derivations for the design of a robust test. Additionally, the a priori probabilities in our work are not required to be equal, which was assumed in all previous works on model mismatch. An example is cognitive radio where the primary user may be idle for most of the time, i.e. $P(\mathcal{H}_0)\gg P(\mathcal{H}_1)$ \cite{cognitive}. Last but not least, the work in this paper allows vector valued observations.\\
The organization of this paper is as follows. In the following section, some background to the minimax optimization problem is given and characterization the saddle value condition is detailed, before the problem definition is stated. Section~\ref{section3} is divided into three parts.
In the first part, the minimax optimization problem is solved and the least favorable distributions, the robust decision rule as well as the robust likelihood ratio, which are later shown to be determined via solving two non-linear equations, are obtained. The second part shows how the problem is simplified if the nominal probability density functions satisfy the symmetry condition. In the third part, the maximum of the robustness parameters, above which a minimax robust test cannot be designed, are derived. In Section~\ref{section4} simulation results that illustrate the validity of the theoretical derivations are detailed. Finally, the paper is concluded in Section~\ref{section5}.

\section{Problem Formulation}
\subsection{Background}
Let $(\Omega,{\mathscr{A}})$ be a measurable space with the probability measures $F_0$, $F_1$, $G_0$, $G_1$, and $G$ on it, having the density functions $f_0$, $f_1$, $g_0$, $g_1$ and $g$ respectively, with respect to some dominating measure $\mu$, i.e., $F_i,G_i,G\ll \mu$, $i\in\{0,1\}$. It is assumed that the nominal measures are distinct, i.e. the condition $F_0= F_1$ $\mu-$almost everywhere is not true. Consider the binary composite hypothesis testing problem
\begin{align}\label{eq1}
\mathcal{H}_0^c&: G= G_0 \nonumber\\
\mathcal{H}_1^c&: G=G_1
\end{align}
where the measures $G_i$ are defined whenever their corresponding density functions $g_i$ belong to the closed ball
\begin{equation}\label{eq7}
{\cal{G}}_i=\{g_i:D(g_i,f_i)\leq \epsilon_i\}\quad i\in\{0,1\},
\end{equation}
where $D$ is a distance between the density functions. In other words, every density function $g_i$ which is at least $\epsilon_i$ close to the nominal density $f_i$ is a member of the uncertainty class ${\mathcal{G}}_i$ and defines $G_i$, $i\in\{0,1\}$. We choose $D$ to be the $\alpha-$divergence i.e.,
\begin{equation}\label{eq89}
D(g,f;\alpha):=\frac{1}{\alpha(1-\alpha)}\left(1-\int_{\Omega}g^\alpha f^{1-\alpha}\mbox{d} \mu\right),\alpha\in\mathbb{R}\backslash \{0,1\}
\end{equation}
since it is a convex distance for every $\alpha$ and it includes various distances as special cases \cite[p.1536]{entropy}.\footnote{Notice that $\alpha-$divergence is preferred against the R\'enyi's $\alpha-$divergence because R\'enyi's $\alpha-$divergence is convex only for $\alpha\in[0,1]$\cite[p.1540]{entropy}}. Given that $y\in\Omega$ has been observed, a randomized decision rule $\delta:\Omega\mapsto [0,1]$ maps each $y$ to a real number in the unit interval. Let $\Delta$ be the set of all decision rules (functions). Then, for any possible choice of $\delta\in \Delta$, the following error types are well defined: first, the false alarm probability
\begin{equation}
P_F(\delta,f_0)=\int_{\Omega}\delta f_0\mbox{d} \mu,\label{eq3}
\end{equation}
second, the miss detection probability
\begin{equation}
P_M(\delta,f_1)=\int_{\Omega}(1-\delta)f_1\mbox{d} \mu\label{eq4},
\end{equation}
and third, the overall error probability
\begin{equation}
P_E(\delta,f_0,f_1)=P(\mathcal{H}_0)P_F(\delta,f_0)+P(\mathcal{H}_1)P_M(\delta,f_1).\label{eq5}
\end{equation}
It is well known that $P_E$ is minimized if the decision rule is chosen to be the likelihood ratio test
\begin{equation}\label{eq6}
\delta(y) = \begin{cases} 0, &l(y)<\rho  \\ \kappa(y), & l(y)=\rho\\ 1, & l(y)> \rho \end{cases},
\end{equation}
where $\rho=P(\mathcal{H}_0)/P(\mathcal{H}_1)$ is some threshold, $l(y):=f_1/f_0(y)$ is the likelihood ratio at observation $y$ and $\kappa:\Omega\rightarrow [0,1]$.
\subsection{Saddle value specification}
In this section, the existence of a saddle value condition due to the functional topology of the minimax optimization problem is shown. Minimax theorem, which is attributed to John von Neumann, gives the necessary conditions such that the existence of a saddle value is guaranteed \cite{ekeland}. However, it is applicable if and only if both sets over which the maximization and minimization is performed are compact. Note that the closed balls $({\mathcal{G}}_0$ and ${\mathcal{G}}_1)$ with respect to the $\alpha-$divergence distance are not compact, therefore Von Neumann's minimax theorem is not applicable in our case. Here, we adopt Sion's minimax theorem \cite{sion},
\begin{align}\label{eq8}
&\sup_{(g_0, g_1)\in {\cal{G}}_0\times{\cal{G}}_1}\min_{\delta\in\Delta}P_E(\delta,g_0,g_1)\nonumber\\
&=\min_{\delta\in\Delta}\sup_{(g_0, g_1)\in {\cal{G}}_0\times{\cal{G}}_1}P_E(\delta,g_0,g_1),
\end{align}
which removes the compactness constraint on the set over which maximization is performed. In order for \eqref{eq8} to be valid the following conditions must hold:
\begin{itemize}
\item The objective function $P_E(\delta,\cdot)$ is real valued, upper semi-continuous and quasi-concave on ${\cal{G}}_0\times{\cal{G}}_1$ for all $\delta\in\Delta$
\item The objective function $P_E(\cdot,(g_0, g_1))$ is lower semi-continuous and quasi-convex on $\Delta$ for all $(g_0, g_1)\in{\mathcal{G}}_0\times{\mathcal{G}}_1$
\item $\Delta$ is a compact convex subset of a linear topological space
\item ${\mathcal{G}}_0\times{\mathcal{G}}_1$ is a convex subset of a linear topological space
\end{itemize}
The first two conditions hold true because $P_E$ is a real valued continuous function, and linear on all three terms $\delta,g_0,g_1$, therefore both convex and concave. The last condition is also true because, all convex combinations of $g_i^0\in{\cal{G}}_i$ and $g_i^1\in{\cal{G}}_i$ are in ${\cal{G}}_i$ since $D$ is a convex distance and the Cartesian product of convex sets is again a convex set.
Similarly, $\Delta$ is a convex set because for any $t\in[0,1]$ and for all $\delta_0,\delta_1\in\Delta$, $t\delta_0+(1-t)\delta_1\in\Delta$. Note that any continuous function is also upper or lower semi-continuous and any convex function is also quasi-convex. Lastly, $\Delta$, which is equivalent to $[0,1]^\Omega$ in infinite dimensional vector space, is the product of uncountably many compact sets $[0,1]$. According to Tychonoff's theorem, $\Delta$ is compact with respect to the product topology \cite{tych1,tych2}. Note that any finitely supported discretization of $g_0$ and $g_1$ makes both ${\mathcal{G}}_0\times {\mathcal{G}}_1$ and $\Delta$ compact with respect to the standard topology. This is a straightforward result of Heine-Borel theorem \cite[Theorem 2.41]{rudin1976}.\\
Accordingly, based on Sions's minimax theorem, there exists a saddle value for the objective function $P_E$, i.e.,
\begin{equation}\label{eq9}
P_E(\delta,\hat{g}_0,\hat{g}_1)\geq P_E(\hat{\delta},\hat{g}_0,\hat{g}_1)\geq P_E(\hat{\delta},g_0,g_1).
\end{equation}
Since $P_E$ is distinct in $g_0$ and $g_1$, we also have
\begin{align}\label{equation12}
&P_F(\hat{\delta},g_0)\leq P_F(\hat{\delta},\hat{g}_0)\nonumber\\
&P_M(\hat{\delta},g_1)\leq P_M(\hat{\delta},\hat{g}_1).
\end{align}
\subsection{Problem definition}
Based on \eqref{equation12}, the minimax optimization problem \eqref{eq8} can be solved considering the Karush-Kuhn-Tucker (KKT) multipliers. Hence, the problem formulation can be restated as
\begin{equation}\label{equation311}
    \begin{aligned}[b]
        \text{Maximization:}\quad &
        \begin{aligned}[t]
            &\hat{g}_0=\mathrm{arg}\sup_{ g_0\in\mathcal{G}_0} P_F(\delta,g_0)\\
             &\quad \text{s.t. $g_0>0$, $\Upsilon(g_0)=\int_{\mathbb{R}}g_0\,\mathrm{d}\mu=1$}\\
            & \hat{g}_1=\mathrm{arg}\sup_{g_1\in\mathcal{G}_1} P_M(\delta,g_1)\\
             &\quad \text{s.t. $g_1>0$, $\Upsilon(g_1)=\int_{\mathbb{R}}g_1\,\mathrm{d}\mu=1$}
        \end{aligned}
        \\[12pt]
        \text{Minimization:}\quad &\hat{\delta}=\mathrm{arg}\min_{\delta\in\Delta} P_E(\delta,\hat{g}_0,\hat{g}_1).
    \end{aligned}
\end{equation}
In \eqref{equation311}, there are two separate maximization problems, which are coupled with the minimization problem through the decision rule ${\delta}$\footnote{In general $\mathrm{arg}\sup$ may not always be achieved since $\mathcal{G}_0$ and $\mathcal{G}_1$ are non-compact sets in the topologies induced by the $\alpha$-divergence distance. In this paper, existence of $\hat{g}_0$ and $\hat{g}_1$ is due to the KKT solution of the minimax optimization problem, which is introduced in Section~\ref{section3}.}

\section{Robust detection with $\alpha-$divergence}\label{section3}
\begin{figure}[ttt]
  \centering
  \psfrag{x}[t][]{$\epsilon=\epsilon_0=\epsilon_1$}
  \psfrag{y}{$y_u$}
  \centerline{\includegraphics[width=8.8cm]{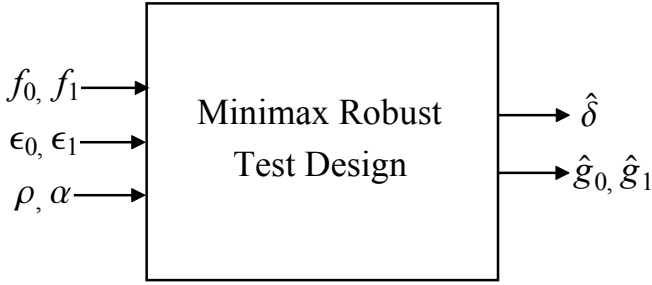}}
\vspace{-3mm}
\caption{Summary of the robust hypothesis testing scheme given by Theorem~\ref{theorem00431}.\hspace{-3mm}\label{fig01}}
\vspace{-2mm}
\end{figure}

\begin{figure}[ttt]
  \centering
  \psfrag{x}[t][]{$\epsilon=\epsilon_0=\epsilon_1$}
  \psfrag{y}{$y_u$}
  \centerline{\includegraphics[width=8.8cm]{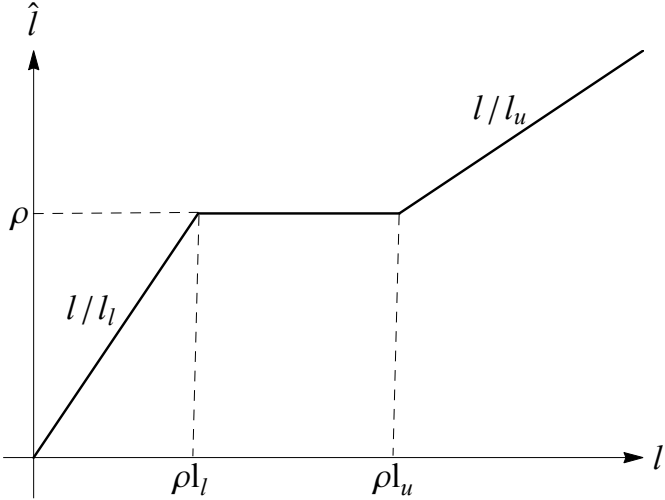}}
\vspace{-3mm}
\caption{Nonlinearity relating the nominal likelihood ratios to the robust likelihood ratios.\hspace{-3mm}\label{fig02}}
\vspace{-2mm}
\end{figure}

The following theorem provides a solution for \eqref{equation311}, which is composed of the least favorable densities $\hat{g}_0$ and $\hat{g}_1$, the robust decision rule $\hat{\delta}$, the robust likelihood ratio function $\hat{l}=\hat{g}_1/\hat{g}_0$ in parametric forms, as well as two non-linear equations from which the parameters can be obtained. Before the statement of the theorem, let $l_l$ and $l_u$ be two real numbers with $0<l_l\leq 1\leq l_u<\infty$. Furthermore, let
\begin{equation}\label{kfunction}
k(l_l,l_u)=\frac{\int_{{\cal{I}}_1}(l-l_l) f_0\mbox{d}\mu}{\int_{{\mathcal{I}}_3}(l_u-l)f_0\mbox{d}\mu},
\end{equation}
\begin{align}\label{zfunction}
&z(l_l,l_u;\alpha,\rho)=\int_{\mathcal{I}_1}f_1\mbox{d}\mu+k(l_l,l_u)\int_{\mathcal{I}_3}f_1\mbox{d}\mu+ \nonumber\\&\hspace{-6mm}\int_{\mathcal{I}_2}\left(\frac{k(l_l,l_u)^{\alpha-1}(l_l^{\alpha-1}-l_u^{\alpha-1})}{l_l^{\alpha-1}-(k(l_l,l_u)l_u)^{\alpha-1}+(k(l_l,l_u)^{\alpha-1}-1)(l/\rho)^{\alpha-1}}\right)^\frac{1}{\alpha-1}f_1\mbox{d}\mu,
\end{align}
where
 \begin{align}\label{integrationdomains}
&{\cal{I}}_1:=\{y:l(y)<\rho{l_l}\} \equiv \{y:\hat{l}(y)<\rho\}\nonumber\\
&{\cal{I}}_2:=\{y:\rho l_l\leq l(y) \leq \rho l_u\} \equiv \{y:\hat{l}(y)=\rho\}\nonumber\\
&{\cal{I}}_3:=\{y:l(y)>\rho{l_u}\} \equiv \{y:\hat{l}(y)>\rho\}
\end{align}
and
\begin{align*}
&\hspace{-2mm}\Phi_1(l,l_l,l_u;\alpha,\rho)=\\
&\hspace{-2mm}\frac{1}{z(l_l,l_u;\alpha,\rho)}\cdot\\
&\left(\frac{k(l_l,l_u)^{\alpha-1}(l_l^{\alpha-1}-l_u^{\alpha-1})}{l_l^{\alpha-1}-(k(l_l,l_u)l_u)^{\alpha-1}+(k(l_l,l_u)^{\alpha-1}-1)(l/\rho)^{\alpha-1}}\right)^\frac{1}{\alpha-1}
\end{align*}
with $\Phi_0 =\Phi_1 l \rho^{-1}$.
\begin{thm}\label{theorem00431}
The least favorable densities
\begin{align}\label{equation00416}
\hat{g}_0 = \begin{cases}  \frac{l_l}{z(l_l,l_u;\alpha,\rho)} f_0, &l<\rho l_l \\
\Phi_0(l,l_l,l_u;\alpha,\rho) f_0,&\rho l_l\leq l\leq \rho l_u \\
\frac{k(l_l,l_u)l_u}{z(l_l,l_u;\alpha,\rho)} f_0, & l> \rho l_u \end{cases},
\end{align}
\begin{align}\label{equation00417}
\hat{g}_1 = \begin{cases} \frac{1}{z(l_l,l_u;\alpha,\rho)} f_1, &l<\rho l_l \\
\Phi_1(l,l_l,l_u;\alpha,\rho) f_1,&\rho l_l\leq l\leq \rho l_u \\
\frac{k(l_l,l_u)}{z(l_l,l_u;\alpha,\rho)} f_1, & l>\rho l_u \end{cases},
\end{align}
and the robust decision rule
\begin{equation}\label{equation00418}
\hspace{-4mm}\hat{\delta}=\begin{cases} 0, &l<\rho l_l  \\
\frac{l_l^{\alpha-1}(l/\rho)^{1-\alpha}-1}{(l_l^{\alpha-1}-(k(l_l,l_u)l_u)^{\alpha-1})(l/\rho)^{1-\alpha}+k(l_l,l_u)^{\alpha-1}-1}, &\rho l_l \leq l\leq \rho l_u\\
1, & l> \rho l_u \end{cases},
\end{equation}
implying the robust likelihood ratio function
\begin{equation}\label{equation00419}
\hat{l}=\frac{\hat{g}_1}{\hat{g}_0} = \begin{cases}l_l^{-1}l, &l<\rho l_l \\
\rho,&\rho l_l\leq l\leq \rho l_u\\
l_u^{-1} l, & l>\rho l_u \end{cases}
\end{equation}
provide a unique solution to \eqref{equation311}. Furthermore, the parameters $l_l$ and $l_u$ can be determined by solving
\begin{align}\label{equation00420}
&\frac{1}{z(l_l,l_u;\alpha,\rho)^\alpha}\Bigg({l_l}^\alpha\int_{\mathcal{I}_1}f_0\mbox{d}\mu+\int_{\mathcal{I}_2} \Phi_0^{'}(l_l,l_u;\alpha,\rho)^\alpha f_0\mbox{d}\mu\nonumber\\
&+(k(l_l,l_u)l_u)^\alpha\int_{\mathcal{I}_3}f_0\mbox{d}\mu\Bigg)=x(\alpha,\varepsilon_0)
\end{align}
and
\begin{align}\label{equation00421}
&\frac{1}{z(l_l,l_u;\alpha,\rho)^\alpha}\Bigg(\int_{\mathcal{I}_1}f_1\mbox{d}\mu+\int_{\mathcal{I}_2} \Phi_1^{'}(l_l,l_u;\alpha,\rho)^\alpha f_1\mbox{d}\mu\nonumber\\
&+k(l_l,l_u)^\alpha\int_{\mathcal{I}_3}f_1\mbox{d}\mu\Bigg)=x(\alpha,\varepsilon_1)
\end{align}
where $\Phi_j^{'}(l_l,l_u;\alpha,\rho)=z(l_l,l_u;\alpha,\rho)\Phi_j$, and $x(\alpha,\varepsilon)=1-\alpha(1-\alpha)\varepsilon$.
\end{thm}
A proof of Theorem~\ref{theorem00431} is given in three stages. In the maximization stage, the Karush-Kuhn-Tucker (KKT) multipliers are used to determine the parametric forms of the LFDs, $\hat{g}_0$ and $\hat{g}_1$, and the robust likelihood ratio function $\hat{l}$. In the minimization stage, the LFDs and the robust decision rule $\hat{\delta}$ are made explicit. Finally, in the optimization stage, four parameters that are needed to design the test are reduced to two parameters without loss of generality.
\begin{proof}
\subsection{Derivation of LFDs and the robust decision rule}
\subsubsection{Maximization step}
Consider the Lagrangian function
\begin{equation}\label{eq14}
L(g_0,\lambda_0,\mu_0)=P_F(\delta,g_0)+\lambda_0(\epsilon_0-D(g_0,f_0;\alpha))+\mu_0(1-\Upsilon(g_0))),
\end{equation}
where $\mu_0$ and $\lambda_0\geq 0$ are the KKT multipliers. It can be seen that $L$ is a  strictly concave functional of $g_0$, as $\partial^2 L/\partial g_0^2<0$ for every $\lambda_0>0$. Therefore, there exists a unique solution to \eqref{eq14}, in case all KKT conditions are met \cite[Chapter 5]{bertsekas2003convex}. More explicitly the Lagrangian can be stated as
\begin{align}\label{eq15}
L(g_0,\lambda_0,\mu_0)&=\int_{\mathbb{R}}\delta g_0-\mu_0 g_0+\frac{\lambda_0}{\alpha(1-\alpha)}\Bigg((1-\alpha)f_0\nonumber\\
&+\alpha g_0-\left(\frac{g_0}{f_0}\right)^\alpha f_0\Bigg)+\lambda_0\epsilon_0+\mu_0 \mbox{d}\mu.
\end{align}
Note that similar to \cite{levy09}, the positivity constraint $g_0\geq 0$ (or $g_1\geq 0$) is not imposed, because for some $\alpha$, this constraint is satisfied automatically, while for others each solution of Lagrangian optimization must be checked for positivity.
To find the maximum of \eqref{eq15}, the directional (G$\hat{\mbox{a}}$teaux's) derivative of the Lagrangian $L$ with respect to $g_0$ in the direction of a function $\psi$ is taken:
\begin{equation}\label{eq16}
\int_{\Omega}\left[\delta-\mu_0+\frac{\lambda_0}{1-\alpha}\left(\left(\frac{g_0}{f_0}\right)^{\alpha-1} -1\right)\right]\psi\mbox{d}\mu.
\end{equation}
Since $\psi$ is arbitrary, $L$ is maximized whenever
\begin{equation}\label{eq17}
\delta-\mu_0+\frac{\lambda_0}{1-\alpha}\left(\left(\frac{g_0}{f_0}\right)^{\alpha-1}-1\right)=0.
\end{equation}
Solving \eqref{eq17} the density function of the LFD $\hat{G}_0$,
\begin{equation}\label{eq18}
\hat{g}_0=\left(\frac{1-\alpha}{\lambda_0}\left(\mu_0-\delta\right)+1\right)^{\frac{1}{\alpha-1}}f_0
\end{equation}
is obtained. Writing the Lagrangian for $P_M$, in a similar way, with the KKT multipliers $\mu_0:=\mu_1$ and $\lambda_0:=\lambda_1$ it follows that
\begin{equation}\label{eq19}
\hat{g}_1=\left(\frac{1-\alpha}{\lambda_1}\left(\mu_1-1+\delta\right)+1\right)^{\frac{1}{\alpha-1}}f_1.
\end{equation}
Accordingly, the robust likelihood ratio function can be obtained as
\begin{equation}\label{eq20}
\hat{l}=\frac{\hat{g}_1}{\hat{g}_0}=\left[\frac{\frac{1-\alpha}{\lambda_1}\left(\mu_1-1+\delta\right)+1}
{\frac{1-\alpha}{\lambda_0}\left(\mu_0-\delta\right)+1}\right]^\frac{1}{\alpha-1}l.
\end{equation}
\subsubsection{Minimization step}
The minimizing decision function is known to be of type \eqref{eq6} with $l$ to be replaced by $\hat{l}$ and $\kappa$ to be determined from \eqref{eq20} via solving $\hat{l}=\rho$ for $\delta:=\hat{\delta}$. For every $\rho$, this results in
\begin{equation}\label{equation38}
\hat{\delta}=\begin{cases} 0, &\hat{l}<\rho  \\
\frac{\lambda_0(-1+\alpha+\lambda_1+\mu_1-\alpha \mu_1)}{(-1+\alpha)(\lambda_0+\lambda_1 (l/\rho)^{1-\alpha})}-\frac{\lambda_1(\lambda_0+\mu_0-\alpha \mu_0)(l/\rho)^{1-\alpha}}{(-1+\alpha)(\lambda_0+\lambda_1 (l/\rho)^{1-\alpha})}, & \hat{l}=\rho.\\
1, & \hat{l}> \rho \end{cases}
\end{equation}
Inserting \eqref{equation38} in \eqref{eq18} and \eqref{eq19}, the least favorable density functions can be obtained as
\begin{equation}\label{equation39}
\hat{g}_0 = \begin{cases} c_1 f_0, &\hat{l}<\rho \\
\Phi_0 f_0,& \hat{l}=\rho \\
c_2 f_0, & \hat{l}> \rho \end{cases},\,\,\,\,\,\,
\hat{g}_1 = \begin{cases} c_3 f_1, &\hat{l}<\rho \\
\Phi_1 f_1,& \hat{l}=\rho \\
c_4 f_1, & \hat{l}> \rho \end{cases},
\end{equation}
where
\begin{equation*}
\hspace{-3mm}c_1=\left(\frac{(1-\alpha)\mu_0+\lambda_0}{\lambda_0}\right)^\frac{1}{\alpha-1}, \,c_2=\left(\frac{(1-\alpha)(\mu_0-1)+\lambda_0}{\lambda_0}\right)^\frac{1}{\alpha-1},
\end{equation*}
\begin{equation*}
\hspace{-3mm}c_3=\left(\frac{(1-\alpha)(\mu_1-1)+\lambda_1}{\lambda_1}\right)^\frac{1}{\alpha-1}, \,c_4=\left(\frac{(1-\alpha)\mu_1+\lambda_1}{\lambda_1}\right)^\frac{1}{\alpha-1}
\end{equation*}
and
\begin{align}
\Phi_0=&\left(\frac{-1+\lambda_0+\lambda_1+\mu_0+\mu_1-\alpha(-1+\mu_0+\mu_1)}{\lambda_0+\lambda_1 (l/\rho)^{1-\alpha}}\right)^\frac{1}{\alpha-1}\label{eq},\\
\Phi_1=&\left(\frac{-1+\lambda_0+\lambda_1+\mu_0+\mu_1-\alpha(-1+\mu_0+\mu_1)}{\lambda_1+\lambda_0 (l/\rho)^{\alpha-1}}\right)^\frac{1}{\alpha-1}.\label{eqeq}
\end{align}
In order to determine the unknown parameters, the constraints in the Lagrangian definition, i.e., $D(\hat{g}_i,f_i,\alpha)=\epsilon_i$ and $\Upsilon(\hat{g}_i)=1$, $i\in\{0,1\}$ are imposed. This leads to \it four \rm non-linear equations:
\begin{align}\label{eq27}
&c_1\int_{\hat{l}<\rho} f_0\mbox{d}\mu+\int_{\hat{l}=\rho}\Phi_0 f_0\mbox{d}\mu+c_2\int_{\hat{l}> \rho} f_0 \mbox{d}\mu=1,\nonumber\\
&c_3\int_{\hat{l}<\rho} f_1\mbox{d}\mu+\int_{\hat{l}=\rho}\Phi_1 f_1\mbox{d}\mu+c_4\int_{\hat{l}> \rho} f_1 \mbox{d}\mu=1,\nonumber\\
&c_1^\alpha\int_{\hat{l}<\rho} f_0\mbox{d}\mu+\int_{\hat{l}=\rho} \Phi_0^\alpha f_0 \mbox{d}\mu+c_2^\alpha\int_{\hat{l}> \rho} f_0\mbox{d}\mu=x(\alpha,\epsilon_0),\nonumber\\
&c_3^\alpha\int_{\hat{l}<\rho} f_1\mbox{d}\mu+\int_{\hat{l}=\rho} \Phi_1^\alpha f_1 \mbox{d}\mu+c_4^\alpha\int_{\hat{l}> \rho} f_1\mbox{d}\mu=x(\alpha,\epsilon_1),
\end{align}
in four parameters, where $x(\alpha,\epsilon)=1-\alpha(1-\alpha)\epsilon$.
\subsubsection{Optimization Step}
In this section, the number of equations as well as the number of parameters are reduced.
This allows the re-definition of $\hat{l}$, $\hat{\delta}$, $\hat{g}_0$ and $\hat{g}_1$ in a more compact form. Let $l_l=c_1/c_3$ and $l_u=c_2/c_4$, then $\hat{l}=\hat{g}_1/\hat{g}_0$ from \eqref{equation39} indicates the equivalence of integration domains, ${\mathcal{I}}_1$, ${\mathcal{I}}_2$ and ${\mathcal{I}}_3$ as defined by \eqref{integrationdomains}.
Applying the following steps in \eqref{eq27}:
\begin{itemize}
\item Consider new domains ${\mathcal{I}}_1$, ${\mathcal{I}}_2$, ${\mathcal{I}}_3$
\item Use the substitutions $c_1:=c_3 l_l$ and $c_2:=c_4 l_u$
\item Divide both sides of the first two equations by $c_3$
\item Equate the resulting equations to each other via $1/c_3$
\end{itemize}
leads to $c_4=k(l_l,l_u)c_3$, where $k(l_l,l_u)$ is as defined by \eqref{kfunction}. Next, the goal is to find a functional $f$ s.t. $\Phi_1=c_3 f(l,l_l,l_u,\alpha)$. Since $\Phi_0 f_0\rho=\Phi_1 f_1$, it follows that $\Phi_0=c_3 f(l,l_l,l_u,\alpha)l\rho^{-1}$, therefore it suffices to evaluate only $\Phi_1$. A step by step derivation of the functional $f$ is given in Appendix~\ref{appendixa}. Accordingly, $\Phi_0$ is also fully specified in terms of the desired parameters and functions. Inserting $\Phi_1$ (which is now a functional of $c_3,l,l_l,l_u,\alpha$), c.f., \eqref{eq37}, into the second equation in \eqref{eq27} and noticing that $c_4=k(l_l,l_u)c_3$ leads to $c_3=1/z(l_l,l_u;\alpha,\rho)$, where $z(l_l,l_u;\alpha,\rho)$ is as defined by \eqref{zfunction}. Applying a similar procedure, which can be found in Appendix~\ref{appendixb}, to $\hat{\delta}$, c.f., \eqref{equation38}, for the case $\hat{l}=\rho$ leads to the robust decision rule $\hat{\delta}$ as given by Theorem~\ref{theorem00431}. The least favorable densities, $\hat{g}_0$ and $\hat{g}_1$, and the robust likelihood ratio function $\hat{l}$ are obtained similarly, by exploiting the connection between the parameters $c_1$, $c_2$, $c_3$, $c_4$ and $l_l$, $l_u$. The same simplifications eventually let the four equations given by \eqref{eq27} to be rewritten as the two equations stated by Theorem~\ref{theorem00431}.
As it was mentioned earlier, both $\hat{g}_0$ and $\hat{g}_1$ are obtained uniquely from the Lagrangian $L$. Hence, $\hat{l}=\hat{g}_1/\hat{g}_0$, and as a result, $\hat{\delta}$ are also unique. It follows that the solution found for \eqref{equation311} by the KKT multipliers approach is unique as claimed.
\end{proof}
Theorem~\ref{theorem00431} can be summarized as illustrated in Figure~\ref{fig01}. In other words, for any choice of pair of nominal density functions $f_0$ and $f_1$, the robustness parameters $\epsilon_0$ and $\epsilon_1$, the Bayesian threshold $\rho$ and the distance parameter $\alpha$, the robust design outputs the least favorable density functions $\hat{g}_0$ and $\hat{g}_1$ and the robust decision rule $\hat{\delta}$.
Notice that $\hat{g}_0$ and $\hat{g}_1$ are the scaled versions (with different scaling factors) of the nominal distributions on $l<\rho l_l$ and $l> \rho l_u$, and in between, they are a composition of both nominals, since $\Phi_0$ and $\Phi_1$ are both functionals of $f_0$ and $f_1$. Interpretation of the decision rule $\hat{\delta}$ is similar, i.e. in the same two regions the robust decision rule is almost surely zero or one, and in between it is a randomized decision rule. The robust version of the nominal likelihood ratio test is a non-linearly transformed version of the nominal likelihood ratios as illustrated by Figure~\ref{fig02}. It is somewhat surprising that the resulting robust likelihood ratio test is the same for the whole family of distances that are parameterized by $\alpha$. In other words, the robust version of the likelihood ratio test, which is given by \eqref{equation00419} is not explicitly a function of $\alpha$.\\
Theorem~\ref{theorem00431} is a generalization of \cite{gul5} in the sense that as $\alpha\rightarrow 1$ and $\rho=1$, the least favorable densities $\hat{g}_0$ and $\hat{g}_1$ as well as the robust decision rule $\hat{\delta}$ reduce to the ones found in \cite{gul5}. The flexibility afforded by the generality of considering a set of distances, called the $\alpha-$divergence, over \cite{gul5} is twofold. First, the designer does not need to search for a suitable distance for modeling errors, and each time test for the applicability to the engineering problem at hand, following tedious steps of derivations. Instead, only the parameter $\alpha$ is required to be determined, which can be done over a training data set via using a suitable search algorithm. Second, the a priori probabilities are not necessarily to be chosen equal. The proposed design with the $\alpha-$divergence covers both cases, in addition to the fact that the choice of the nominal probability distributions also does not require any assumption. Additional constraints on the choice of nominal distributions as well as on the robustness parameters simplify the design as introduced in the next section.
\subsection{Simplified model with additional constraints}
In some cases, evidence that the following assumption holds may be available:
{\begin{assumption}
The nominal likelihood ratio $l$ is monotone and the nominal density functions are symmetric, i.e., $f_1(y)=f_0(-y)\,\forall y$
\end{assumption}}If, additionally, the robustness parameters are set to be equal, $\epsilon=\epsilon_0=\epsilon_1$, or in other words $x(\alpha,\epsilon)=x(\alpha,\epsilon_0)=x(\alpha,\epsilon_1)$, it follows that
\begin{align}\label{equation29}
\delta(y)& =1-\delta(-y)\nonumber\\
& \makebox*{${}={}$}{$ \Uparrow $} \nonumber\\
\begin{rcases}\begin{cases*} l_u =1/l_l \\y_u =-y_l \end{cases*}\end{rcases}\Longleftrightarrow\begin{cases*} c_2\\c_1\end{cases*} & \begin{rcases} = c_3\\ = c_4 \end{rcases}
\Longleftrightarrow\begin{rcases}\begin{cases*}\lambda_0=\lambda_1\\
 \mu_0=\mu_1\end{cases*}\end{rcases} \nonumber\\
&\makebox*{${}={}$} {$\Updownarrow$} \nonumber\\
g_1(y)& =g_0(-y)
\end{align}
where $y_l=l^{-1}(l_l)$ and $y_u=l^{-1}(l_u)$. These relationships are straightforward and therefore the proofs are omitted. Notice that, due to monotonicity of $l$, the limits of integrals $\mathcal{I}_1$, $\mathcal{I}_2$ and $\mathcal{I}_3$ should be re-arranged e.g.,
\begin{align*}
{\mathcal{I}}_1:&=\{y:l(y)<\rho{l_l}\}\nonumber\\
&\equiv \{y: y<l^{-1}(\rho l(y_l))\}\equiv \{y: y<l^{-1}(\rho l(-y_u))\}.
\end{align*}
The symmetry assumption implies:
\begin{align}\label{equation28}
x(\alpha,\epsilon)&=\int_{\mathbb{R}}\left(\frac{g_1(y)}{f_1(y)}\right)^\alpha f_1(y)\mbox{d}y=\int_{\mathbb{R}}\left(\frac{g_1(y)}{f_0(-y)}\right)^\alpha f_0(-y)\mbox{d}y\nonumber\\
&=\int_{\mathbb{R}}\left(\frac{g_0(y)}{f_0(y)}\right)^\alpha f_0(y)\mbox{d}y=\int_{\mathbb{R}}\left(\frac{g_0(-y)}{f_0(-y)}\right)^\alpha f_0(-y)\mbox{d}y\nonumber\\
&=\int_{\mathbb{R}}\left(\frac{g_0(-y)}{f_1(y)}\right)^\alpha f_1(y)\mbox{d}y
\end{align}
for all $\alpha$ and $\epsilon$ and, it also implies $l(y)=1/l(-y)$ and as a result $\hat{l}(y)=1/\hat{l}(-y)$ for all $y$. Hence,  $g_1(y)=g_0(-y)\,\forall y$ is a solution and all the simplifications in \eqref{equation29} follow. This reduces the four equations given by \eqref{eq27} to two:
\begin{align}\label{equation27}
c_4&=\Bigg(l(y_u)\int_{-\infty}^{y_l^*}f_1(y)\mbox{d}y\nonumber\\
&+\int_{y_l^*}^{y_u^*}\left(\frac{1+l(y_u)^{\alpha-1}}{1+(l(y)/\rho)^{\alpha-1}}\right)^{\frac{1}{\alpha-1}}f_1(y)\mbox{d}y\nonumber\\
&+\int_{y_u^*}^\infty f_1(y)\mbox{d}y\Bigg)^{-1}
\end{align}
and
\begin{align}\label{equation26}
&{c_4}^\alpha\Bigg(l(y_u)^\alpha\int_{-\infty}^{y_l^*}f_1(y)\mbox{d}y\nonumber\\
&+\int_{y_l^*}^{y_u^*}\left(\frac{1+l(y_u)^{\alpha-1}}{1+(l(y)/\rho)^{\alpha-1}}\right)^{\frac{\alpha}{\alpha-1}}f_1(y)\mbox{d}y\nonumber\\
&+\int_{y_u^*}^\infty f_1(y)\mbox{d}y\Bigg)=x(\alpha,\epsilon),
\end{align}
where $y_l^*(y_u)=l^{-1}(\rho l(-y_u))$ and $y_u^*(y_u)=l^{-1}(\rho l(y_u))$. These two equations can then be combined into a single equation
\begin{align}\label{equation25}
&l(y_u)^\alpha\int_{-\infty}^{y_l^*}f_1(y)\mbox{d}y+\int_{y_l^*}^{y_u^*}\left(\frac{1+l(y_u)^{\alpha-1}}{1+(l(y)/\rho)^{\alpha-1}}\right)^{\frac{\alpha}{\alpha-1}}f_1(y)\mbox{d}y  \nonumber\\
&+\int_{y_u^*}^\infty f_1(y)\mbox{d}y-x(\alpha,\epsilon)\Bigg(l(y_u)\int_{-\infty}^{y_l^*}f_1(y)\mbox{d}y\nonumber\\
&+\int_{y_l^*}^{y_u^*}\left(\frac{1+l(y_u)^{\alpha-1}}{1+(l(y)/\rho)^{\alpha-1}}\right)^{\frac{1}{\alpha-1}}f_1(y)\mbox{d}y+\int_{y_u^*}^\infty f_1(y)\mbox{d}y\Bigg)^\alpha=0,
\end{align}
from where the parameter $y_u$ can easily be determined. Obviously, the computational complexity is reduced considerably with the aforementioned assumptions, i.e., when \eqref{equation25} is
compared to \eqref{equation00420} and \eqref{equation00421}. Note that when $\rho=1$, we have $y_l^*=-y_u$ and $y_u^*=y_u$ and if additionally $\alpha\rightarrow 1$, \eqref{equation25} reduces to \cite{levy09}, cf.~\cite{entropy}.

\subsection{Limiting Robustness Parameters}\label{limit}
The existence of a minimax robust test strictly depends on the pre-condition that the uncertainty sets $\mathcal{G}_i$ are distinct. To satisfy this condition, Huber suggested $\epsilon_i$ to be chosen small, see \cite[p.3]{hube65}. Dabak \cite{dabak} does not mention how to choose the parameters, whereas Levy gives an implicit bound as the relative entropy between the half way density $f_{1/2}=f_0^{1/2}f_1^{1/2}/z$ and the nominal density $f_0$, i.e., $\epsilon<D(f_{1/2},f_0)$, where $z$ is a normalizing constant. In the sequel, we show explicitly which pairs of parameters $(\epsilon_0,\epsilon_1)$ are valid to design a minimax robust test for the $\alpha-$divergence distance.\\
The limiting condition for the uncertainty sets to be disjoint is $\hat{G}_0=\hat{G}_1$ $\mu$-a.e. It is clear from the saddle value condition \eqref{eq} that for any possible choice of $(\epsilon_0,\epsilon_1)$, which results in $\hat{G}_0=\hat{G}_1$, it is true that $P_E\leq 1/2$ for all $(g_0\times g_1)\in\mathcal{G}_0\times\mathcal{G}_1$. Since infinitesimally smaller parameters guarantee the strict inequality $P_E< 1/2$, it is sufficient to determine all possible pairs which result in $\hat{G}_0=\hat{G}_1$. A careful inspection suggests that the LFDs are identical whenever $l_l\rightarrow \inf l$ and $l_u\rightarrow \sup l$. For this choice $\mathcal{I}_1$ and $\mathcal{I}_3$ are empty sets and the density functions under each hypothesis are defined only on $\mathcal{I}_2$. Without loss of generality, assume that $\alpha<1$, $\inf l=0$ and $\sup l=\infty$.
For this choice $l_l\rightarrow 0$ implies $\mu_1=\lambda_1/(\alpha-1)+1$ and $l_u\rightarrow \infty$ implies $\mu_0=\lambda_0/(\alpha-1)+1$. Inserting these into one of the first two
equations in \eqref{eq27}, gives
\begin{equation}\label{equation24}
\int_{\Omega}\left(\lambda_0 f_0(y)^{1-\alpha}+\lambda_1\rho^{\alpha-1} f_1(y)^{1-\alpha}\right)^{\frac{1}{1-\alpha}}\mbox{d}y=\left({1-\alpha}\right)^{\frac{1}{1-\alpha}}.
\end{equation}
Similarly, from the third and fourth equations it follows that
\begin{align}\label{equation23}
&\int_{\mathbb{R}}\left(\lambda_0 f_0(y)^{\frac{1-\alpha}{\alpha}}+\lambda_1 \rho^{\alpha-1} f_1(y)^{1-\alpha}f_0(y)^{\frac{(\alpha-1)^2}{\alpha}}\right)^{\frac{\alpha}{1-\alpha}}\mbox{d}y\nonumber\\
&=\left({1-\alpha}\right)^{\frac{\alpha}{1-\alpha}}x(\alpha,\epsilon_0)
\end{align}
and
\begin{align}\label{equation22}
&\int_{\mathbb{R}}\left(\lambda_1 f_1(y)^{\frac{1-\alpha}{\alpha}}+\lambda_0 \rho^{1-\alpha} f_0(y)^{1-\alpha}f_1(y)^{\frac{(\alpha-1)^2}{\alpha}}\right)^{\frac{\alpha}{1-\alpha}}\mbox{d}y\nonumber\\
&=\left({1-\alpha}\right)^{\frac{\alpha}{1-\alpha}}x(\alpha,\epsilon_1).
\end{align}
Given $\rho$ and $\alpha$, \eqref{equation24}, \eqref{equation23}, and \eqref{equation22} can jointly be solved to determine the space of maximum robustness parameters. As an example, consider $\Omega=\mathbb{R}$, $\rho=1$ and $\alpha=1/2$. This choice of $\alpha$ corresponds to the squared Hellinger distance with an additional scaling factor of $1/\alpha(1-\alpha)=4$. Let $a=\int_{-\infty}^{\infty}\sqrt{f_0(y)f_1(y)}\mbox{d}y$. Then, the Equations \eqref{equation24}-\eqref{equation22} reduce to the polynomials in the Lagrangian multipliers $\lambda_0$ and $\lambda_1$,
\begin{equation}\label{equation21}
\lambda_0^2+\lambda_1^2+2\lambda_0\lambda_1 a-\frac{1}{4}=0,
\end{equation}
\begin{equation}\label{equation20}
4-8\lambda_0-8\lambda_1 a-\epsilon_0=0,
\end{equation}
\begin{equation}\label{equation19}
4-8\lambda_1-8\lambda_0 a-\epsilon_1=0,
\end{equation}
respectively. Solving \eqref{equation20} and \eqref{equation19} for $\lambda_0$ and $\lambda_1$, respectively, and inserting the results into Equation \eqref{equation21} we get
\begin{equation}\label{equation18}
2 \epsilon_1  (a (\epsilon_0-4)+4)-(4 a+\epsilon_0-4)^2-\epsilon_1^2=0.
\end{equation}
Equation \eqref{equation18} is quadratic in $a$ and has two roots. One of the roots results in $a=1$ for all $\epsilon_0=\epsilon_1$, which is not plausible. Therefore, the correct root is,
\begin{equation}\label{equation17}
a=\frac{1}{16} \left(16-4 \epsilon_1+\epsilon_0 (\epsilon_1-4)-\sqrt{(\epsilon_0-8) \epsilon_0 (\epsilon_1-8) \epsilon_1}\right).
\end{equation}
Notice that \eqref{equation17} is symmetric in $\epsilon_0$ and $\epsilon_1$, i.e., $a(\epsilon_0,\epsilon_1)=a(\epsilon_1,\epsilon_0)$ for all $(\epsilon_0,\epsilon_1)$, as expected. Since $0\leq a\leq 1$ is known a priori, given a choice of $\epsilon_i$, the corresponding $\epsilon_{1-i}$ can be determined from \eqref{equation17} easily, c.f., Section~\ref{section4}. A special case occurs whenever $\epsilon=\epsilon_0=\epsilon_1$, which simplifies \eqref{equation17} to
\begin{equation}\label{equation16}
\epsilon_{{\max}}=4-2\sqrt{2(1+a)}.
\end{equation}
Maximum robustness parameters given by \eqref{equation17} and \eqref{equation16} are in agreement with the ones found in \cite{gul3}. The case $\alpha>1$, which implies $\mu_0=\lambda_0/(\alpha-1)$ and $\mu_1=\lambda_1/(\alpha-1)$, can be examined similarly.

\section{Simulations}\label{section4}
\begin{figure}[thb]
  \centering
  \psfrag{x}[t][]{$\epsilon=\epsilon_0=\epsilon_1$}
  \psfrag{y}{$y_u$}
  \centerline{\includegraphics[width=8.8cm]{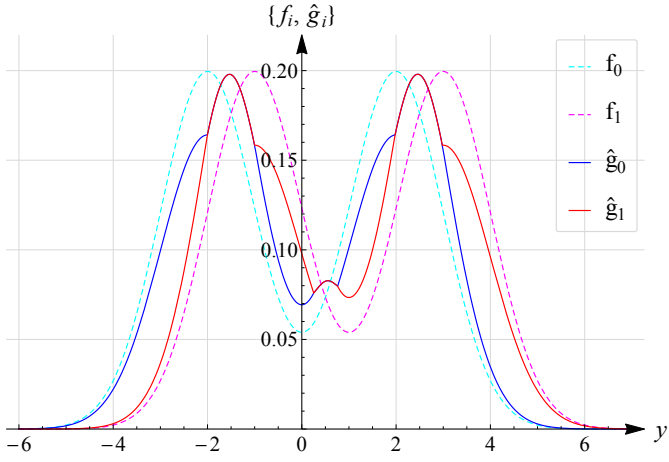}}
\vspace{-3mm}
\caption{Nominal densities and the corresponding least favorable densities for $\rho=1$, $\alpha=4$, $\epsilon_0=0.02$ and $\epsilon_1=0.03$.\hspace{-3mm}\label{fig2}}
\vspace{-2mm}
\end{figure}

\begin{figure}[thb]
  \centering
  \psfrag{x}[t][]{$\epsilon=\epsilon_0=\epsilon_1$}
  \psfrag{y}{$y_u$}
  \centerline{\includegraphics[width=8.8cm]{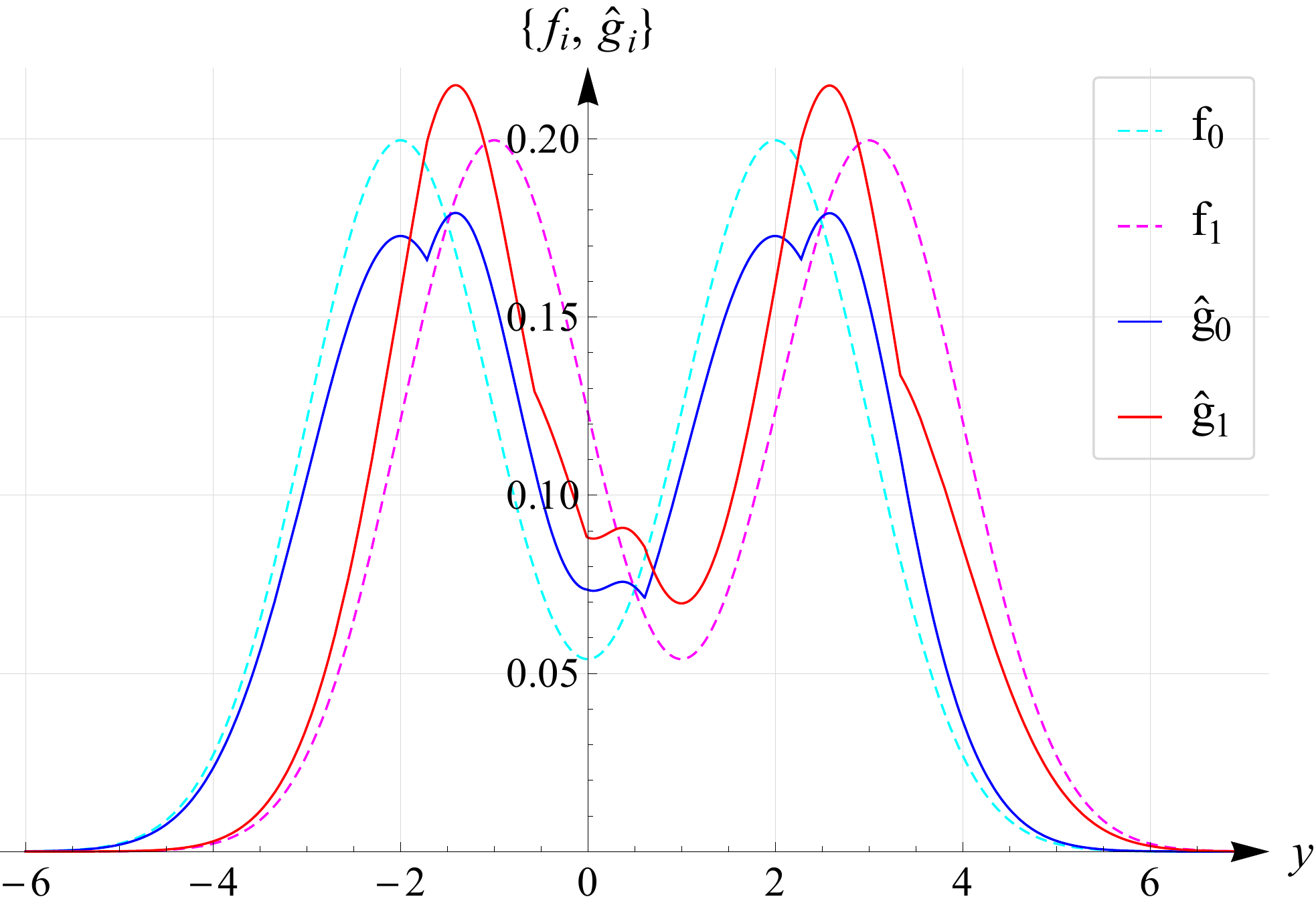}}
\vspace{-3mm}
\caption{Nominal densities and the corresponding least favorable densities for $\rho=1.2$, $\alpha=4$, $\epsilon_0=0.02$ and $\epsilon_1=0.03$.\hspace{-3mm}\label{fig3}}
\vspace{-2mm}
\end{figure}

\begin{figure}[thb]
  \centering
  \psfrag{x}[t][]{$\epsilon=\epsilon_0=\epsilon_1$}
  \psfrag{y}{$y_u$}
  \centerline{\includegraphics[width=8.8cm]{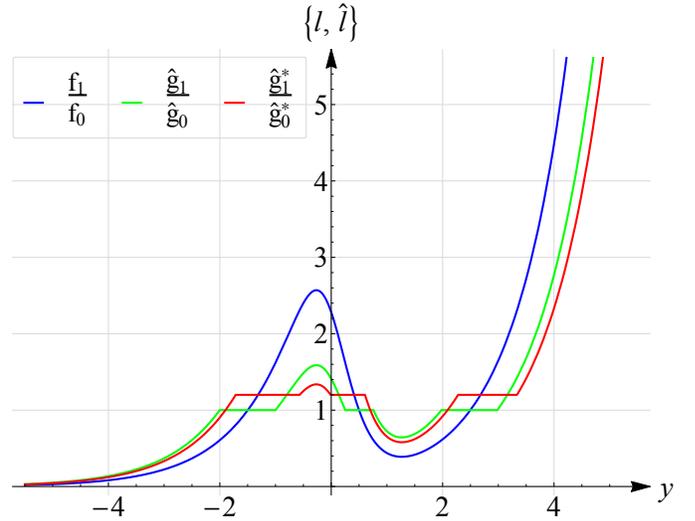}}
\vspace{-3mm}
\caption{Nominal and least favorable likelihood ratios ($\hat{g}_1/\hat{g}_0$ for $\rho=1$ and $\hat{g}_1^*/\hat{g}_0^*$ for $\rho=1.2$) for $\alpha=4$, $\epsilon_0=0.02$ and $\epsilon_1=0.03$. \hspace{-3mm}\label{fig4}}
\vspace{-2mm}
\end{figure}

\begin{figure}[thb]
  \centering
  \psfrag{x}[t][]{$\epsilon=\epsilon_0=\epsilon_1$}
  \psfrag{y}{$y_u$}
  \centerline{\includegraphics[width=8.8cm]{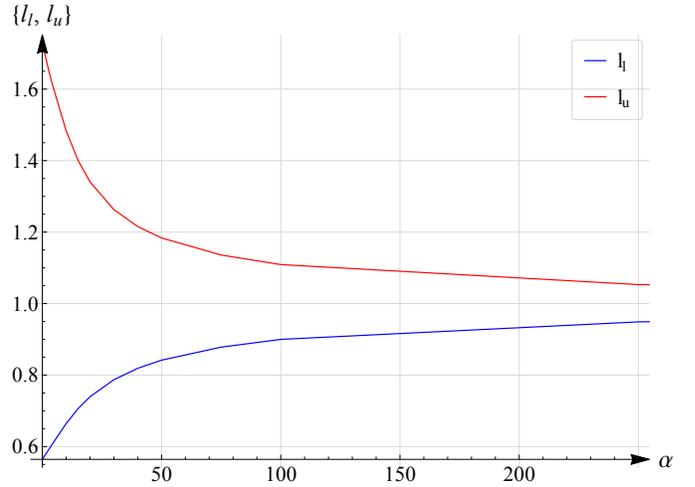}}
\vspace{-3mm}
\caption{Lower and upper thresholds, $l_l$ and $l_u$, for a variable $\alpha$, $\rho=1$, $\epsilon_0=0.02$ and $\epsilon_1=0.03$.\hspace{-3mm}\label{fig5}}
\vspace{-2mm}
\end{figure}

\begin{figure}[thb]
  \centering
  \psfrag{x}[t][]{$\epsilon=\epsilon_0=\epsilon_1$}
  \psfrag{y}{$y_u$}
  \centerline{\includegraphics[width=8.8cm]{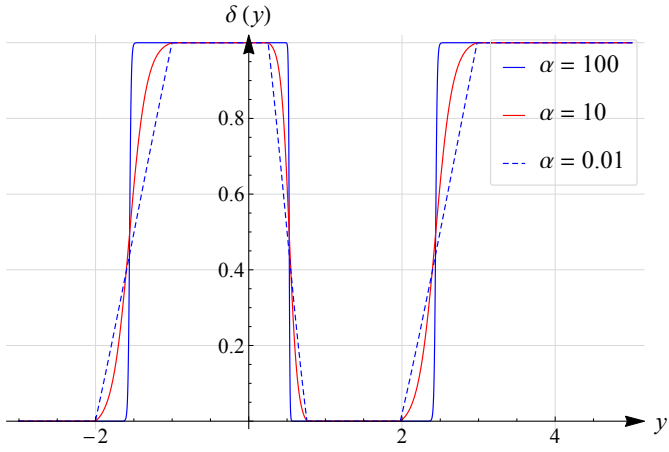}}
\vspace{-3mm}
\caption{The decision rule $\hat{\delta}$ for $\alpha=\{0.01,10,100\}$, $\rho=1$, $\epsilon_0=0.02$ and $\epsilon_1=0.03$.\hspace{-3mm}\label{fig6}}
\vspace{-2mm}
\end{figure}

\begin{figure}[thb]
  \centering
  \psfrag{x}[t][]{$\epsilon=\epsilon_0=\epsilon_1$}
  \psfrag{y}{$y_u$}
  \centerline{\includegraphics[width=8.8cm]{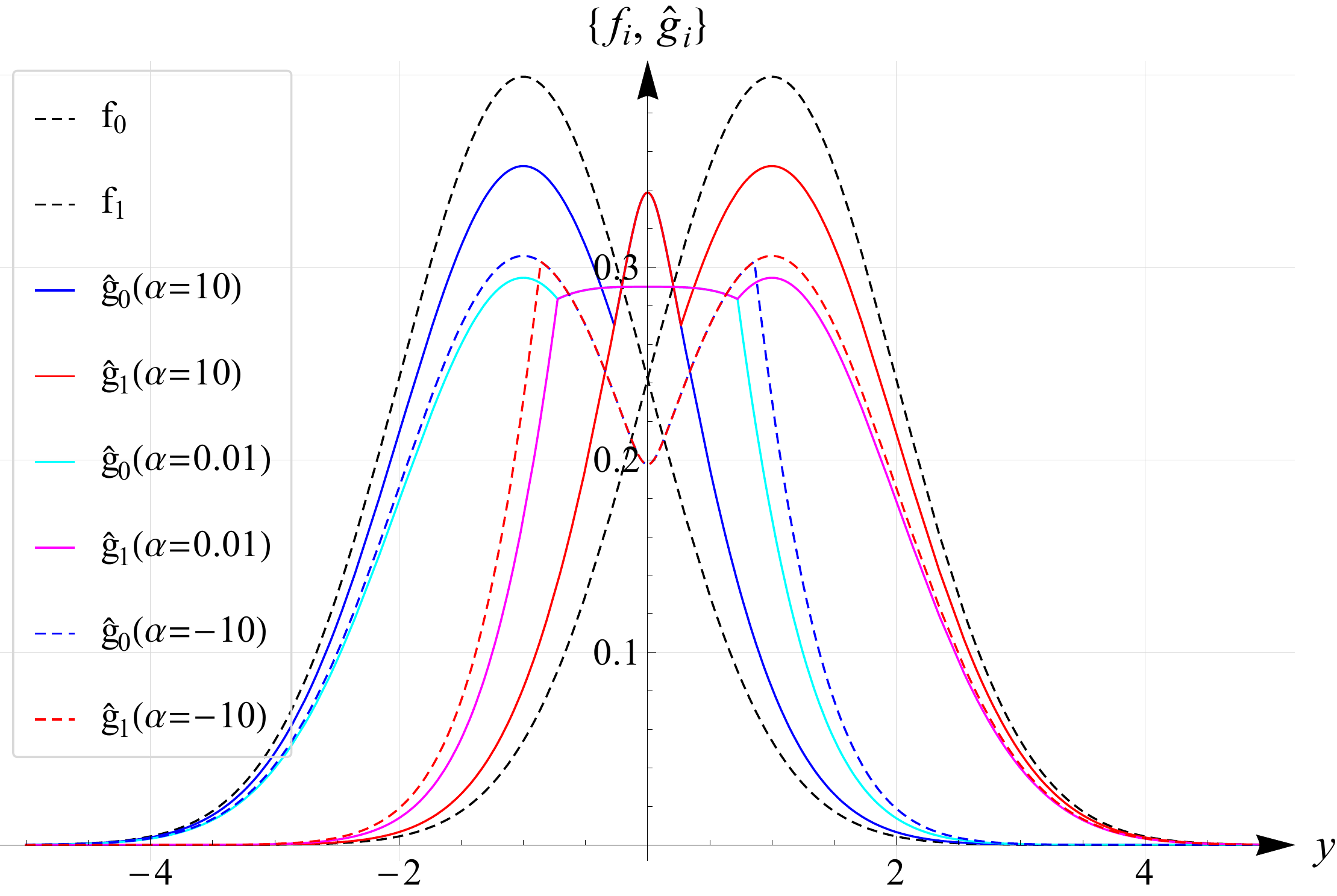}}
\vspace{-3mm}
\caption{Nominal densities and the corresponding least favorable densities for $\rho=1$, $\epsilon_0=0.1$ and $\epsilon_1=0.1$.\hspace{-3mm}\label{fig7}}
\vspace{-2mm}
\end{figure}

\begin{figure}[thb]
  \centering
  \psfrag{x}[t][]{$\epsilon=\epsilon_0=\epsilon_1$}
  \psfrag{y}{$y_u$}
  \centerline{\includegraphics[width=8.8cm]{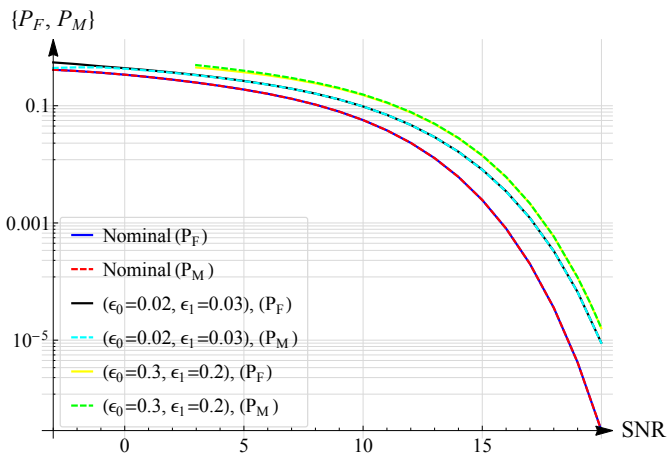}}
\vspace{-3mm}
\caption{False alarm and miss detection probabilities of $\delta$, \eqref{eq7}, $(\rho=1)$ for $(f_0,f_1)$ compared to that of the robust decision rule $\hat{\delta}$ for $(\hat{g}_0,\hat{g}_1)$ when SNR is varied.\hspace{-3mm}\label{fig8}}
\vspace{-2mm}
\end{figure}

\begin{figure}[thb]
  \centering
  \psfrag{x}[t][]{$\epsilon=\epsilon_0=\epsilon_1$}
  \psfrag{y}{$y_u$}
  \centerline{\includegraphics[width=8.8cm]{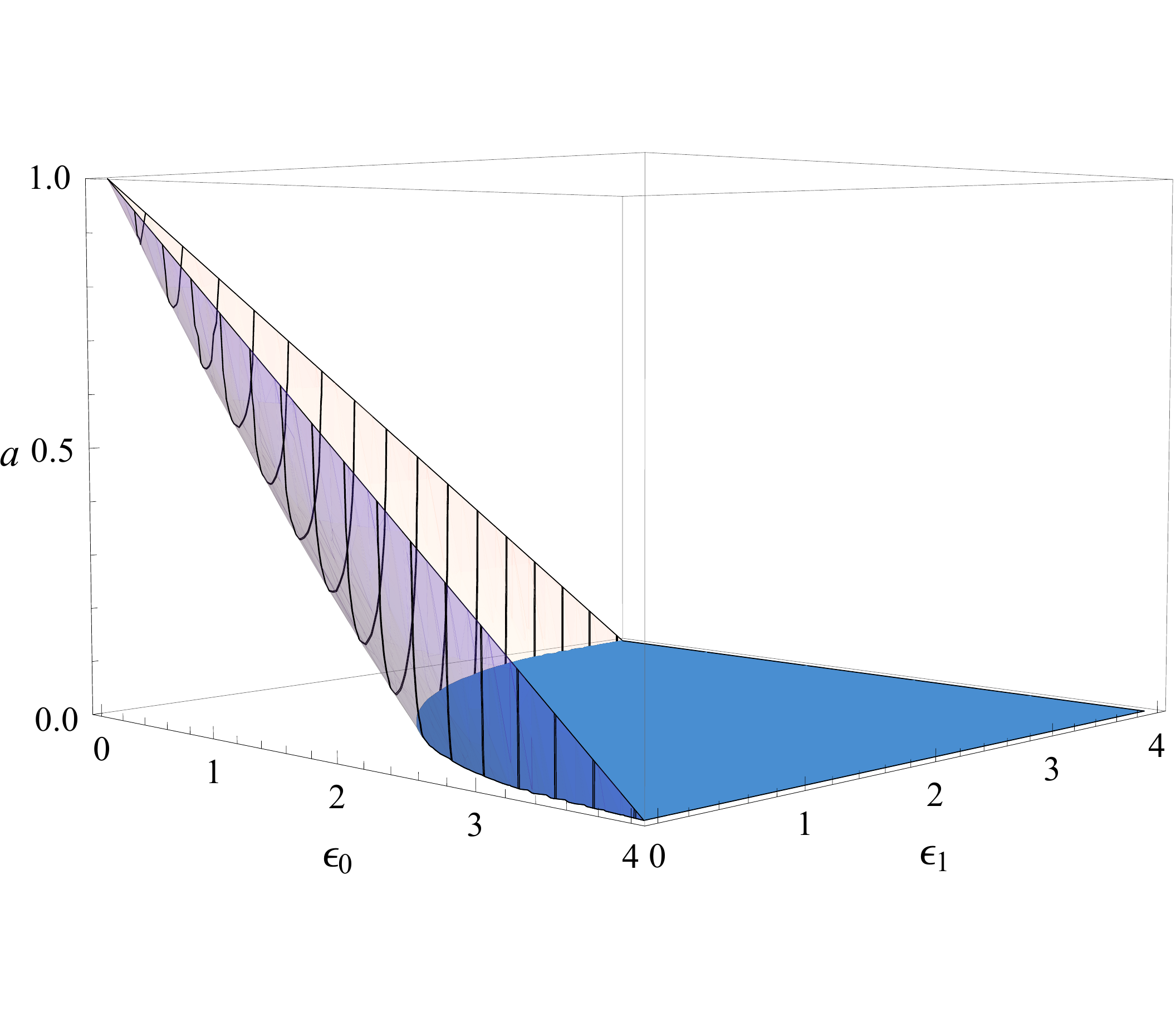}}
\vspace{-3mm}
\caption{All allowable pairs of maximum robustness parameters, $(\epsilon_0,\epsilon_1)$, w.r.t. all distances $a\in[0,1]$  for $\alpha=0.5$.\hspace{-3mm}\label{fig1}}
\vspace{-2mm}
\end{figure}

In this section, some simulations are performed to illustrate the theoretical derivations. Consider a simple hypothesis testing problem
\begin{align}\label{equation15}
\mathcal{H}_0^s&: Y=W \nonumber\\
\mathcal{H}_1^s&: Y=W+A
\end{align}
where $A>0$ is a known DC signal, $W$ is a random variable which follows a symmetric Gaussian mixture distribution
\begin{equation}
W\sim \frac{1}{2}\left(\mathcal{N}(-\mu,\sigma^2)+\mathcal{N}(\mu,\sigma^2)\right),
\end{equation}
where $\mathcal{N}(\mu,\sigma^2)$ is a Gaussian distribution with mean $\mu$ and variance $\sigma^2$ and $Y$ is a random variable on $\Omega=\mathbb{R}$, which is consistent with the data sample $y$. To account for uncertainties on $Y$ under both hypotheses, let
\begin{align}
F_0(y):=&P(Y<y|\mathcal{H}_0^s)\quad\forall y\nonumber\\
F_1(y):=&P(Y<y|\mathcal{H}_1^s)\quad\forall y
\end{align}
be the nominal distributions, having the density functions $f_0$ and $f_1$ for the binary composite hypothesis testing problem given by \eqref{eq1} and \eqref{eq7}. Note that the symmetry condition, $f_1(y)= f_0(-y)$ for all $y$, does not hold, and $l=f_1/f_0$ is not monotone.
Assume $\mu=2$, $\sigma=1$ and $A=1$ and let the robustness parameters be $\epsilon_0=0.02$ and $\epsilon_1=0.03$ for the $(\alpha=4)-$divergence distance.
This example demonstrates an extreme case, for which no straightforward simplification to the equations \eqref{equation00420} and \eqref{equation00421} exists,
both in terms of reducing the number of equations as well as for the domain of integrals. Figure~\ref{fig2} illustrates the nominal density functions $f_0$ and $f_1$
along with the density functions of the corresponding least favorable densities (LFD)s $g_0$ and $g_1$, for an equal a priori probability $\rho=1$.
It can be observed that LFDs intersect in three distinct intervals, each at the neighborhood of $y=-1.5+j$ for $j\in\{0,2,4\}$.
In Fig.~\ref{fig3}, the same simulation is repeated for $\rho=1.2$. In Fig.~\ref{fig4} the nominal and least favorable likelihood ratios for the same example are shown. As it was given by \eqref{equation00419},
robustification of the simple hypothesis test corresponds to a non-linear transformation of the nominal likelihood ratios.\\
In the next simulation, all the parameters are fixed as before, except for $\alpha$. We are especially interested in the change in the lower and upper thresholds, $l_l$ and $l_u$, for varying $\alpha$.
Figure~\ref{fig5} illustrates the outcome of this simulation for $\rho=1$. We can see that $l_l$ and $l_u$ tend to $1$ for $\alpha\rightarrow\infty$. It is not straightforward to derive this from \eqref{equation00420} and \eqref{equation00421} for any $f_0$ and $f_1$. However, if there exists a solution, which is true and unique by the KKT multipliers approach, it should satisfy $D(f,g;\alpha)=\epsilon_i$ for any $\alpha>0$ and for all allowable $\epsilon_i$, cf. Section~\ref{limit}. Assume that $g$ is fixed and it does not depend on $\alpha$. Then, the integral $\int_\mathbb{R} g^\alpha f^{1-\alpha}\mbox{d}\mu$ is $1$ at $\alpha=0$ and $\alpha=1$, convex in $\alpha$, and it is positive for all $\alpha>0$, $f$ and $g$. Hence, $\lim_{\alpha\rightarrow \infty}\int_\mathbb{R} g^\alpha f^{1-\alpha}\mbox{d}\mu=\infty$ and $\lim_{\alpha\rightarrow \infty}D(f,g;\alpha)$ is indeterminate. Using L'Hospital's rule twice we obtain
\begin{equation}
\hspace{-2mm}K=\lim_{{\alpha\rightarrow\infty}}D(g,f;\alpha)=\lim_{{\alpha\rightarrow\infty}}\frac{\int_\mathbb{R} \log^2(g/f)(g/f)^\alpha f \mathrm{d}\mu}{2}.
\end{equation}
The integral $\int_\mathbb{R} \log^2(g/f)(g/f)^\alpha f \mathrm{d}\mu$ is also positive and convex in $\alpha$. This implies $K\rightarrow\infty$ for $\alpha\rightarrow\infty$. Now, assume that $g$ depends on $\alpha$ and tends to a limiting distribution $g^*$ for $||g^*-f||>0$, when $\alpha\rightarrow\infty$. Then, our conclusion does not change, i.e., $K\rightarrow\infty$ for $\alpha\rightarrow\infty$. Since $D(f,g;\alpha)$ is finite, we require that $\alpha\rightarrow\infty\Longrightarrow g^*\rightarrow f$. Consequently, from \eqref{equation00416} and \eqref{equation00417}, $\hat{g}_i\rightarrow f_i$ whenever $l_l\rightarrow 1$ and $l_u\rightarrow 1$ explains the asymptotic of Figure~\ref{fig5} for any pair $(f_0,f_1)$.\\
Based on simulation results the following are conjectured:
\begin{itemize}
\item For a fixed $\epsilon_0$ and $\epsilon_1$, increasing $\alpha$ leads to a monotone decrease in $l_u$ and monotone increase in $l_l$ on $\mathbb{R}^+\backslash \{0,1\}$.
\item For a fixed $\alpha$, increasing $\epsilon_0$, $\epsilon_1$ or both introduces a non-decrease in $l_u$, non-increase in $l_l$, or both, given that $\epsilon_0$ and $\epsilon_1$ are less than their allowable maximum, cf. Section~\ref{limit}.
\end{itemize}
The proof of these conjectures is an open problem.\\
From \eqref{equation00420} and \eqref{equation00421}, it is clear that given a pair $(\epsilon_0,\epsilon_1)$, a slight change in $\alpha$ changes the equations completely and in general $l_l$ and $l_u$ are functions of $\alpha$. In Figure~\ref{fig6}, the robust decision rule $\hat{\delta}$ for various $\alpha$ values is plotted, without considering the dependency of $l_l$ and $l_u$ on $\alpha$. To do this, $l_l\approx 0.605$ and $l_u\approx 1.618$, that are found for $\rho=1$, $\alpha=4$, $\epsilon_0=0.02$ and $\epsilon_0=0.03$, are fixed constants in \eqref{equation00418}. Then, for $\alpha=\{0.01,10,100\}$, \eqref{equation00418} is plotted. The decision rule $\hat{\delta}$ tends to a step like function for an increasing $\alpha$, whereas for a smaller $\alpha$, i.e., $\alpha=0.01$, the decision rule is almost linear at the domain of the likelihood ratio for which $\hat{l}=1$.
This result is also in agreement with the previous findings; $\hat{\delta}$ tends to a non-randomized likelihood ratio test for $\alpha\rightarrow \infty$, for which we obtained $\hat{g}_i\rightarrow f_i$ and for $(f_0,f_1)$ optimum decision rule is
known to be a non-randomized likelihood ratio test.\\
In the following simulation, the simplified model ($f_0(y)=f_1(-y)$) is tested for mean shifted Gaussian distributions; $F_0\sim\mathcal{N}(\mu_0,\sigma^2)$ and $F_1\sim\mathcal{N}(\mu_1,\sigma^2)$ with means $\mu_0=-1$, $\mu_1=1$ and variance $\sigma^2=1$. The parameters of the composite test are chosen to be $\rho=1$, $\epsilon_0=0.1$ and $\epsilon_1=0.1$.
Here, our main interest is to observe the change in overlapping regions of least favorable density pairs for various $\alpha$. Figure~\ref{fig7} illustrates the outcome of this simulation.
It can be seen that the overlapping region is convex for a negative $\alpha$, ($\alpha=-10$) almost constant for $\alpha=0.01$ and concave for a positive $\alpha$, ($\alpha=10$).
For the sake of clarity only three examples of $\alpha$ are plotted.\\
In Figure~\ref{fig8}, the false alarm and miss detection probabilities of the likelihood ratio test $\delta$ for $(f_0,f_1)$ are graphed and compared with the robust test $\hat{\delta}$ for $(\hat{g}_0,\hat{g}_1)$. Two different robust parameter pairs and various signal to noise ratios (SNR)s, i.e., $\mbox{SNR}=20\log({A}/{\sigma})$ are considered. It can be seen that increasing the robustness parameters increases the false alarm and miss detection probabilities for all SNRs, as expected. The difference between false alarm and miss detection probabilities for the same robust test is small and it is more pronounced for low SNRs. For high SNRs the performance of two robust tests are close to each other. The reason is that for high SNRs maximum allowable robustness parameters become relatively high compared to the parameters of both robust settings.
Although the nominal test has the lowest error rates, its performance can degrade considerably under uncertainties in the nominal model. The robust tests, on the other hand,
have slightly higher error rates, but guaranteed power of the test, which indicates the trade-off between performance and robustness.
Finally, in the last simulation, the 3D boundary surface of the maximum robustness parameters is determined for $\alpha=0.5$ \eqref{equation17} and is shown in Figure~\ref{fig1}. This surface has a cropped rotated cone like shape, which is symmetric about its main diagonal, i.e.,
with respect to the plane $\epsilon_0=\epsilon_1$ on the space $(\epsilon_0,\epsilon_1,a)$. Notice that except for the points on the cone like shape that intersect with the
$(\epsilon_0,\epsilon_1,a=0)$ plane, all other points on $(\epsilon_0,\epsilon_1,a=0)$ that are plotted in blue color are un-defined (rather than being valid points with $a=0$), implying that for those points no minimax robust test exists.


\section{Conclusion}\label{section5}
A robust version of the likelihood ratio test considering $\alpha-$divergence as the distance to characterize the uncertainty sets has been proposed.
The existence of a saddle value to the minimax optimization problem was shown by adopting Sion's minimax theorem.
The least favorable distributions, the robust decision rule as well as the robust version of the likelihood ratio test were derived in two
parameters and in three distinct regions on the co-domain of the nominal likelihood ratio.
Two equations from where the parameters can be determined were also derived.
It was found that the robust likelihood ratio doesn't depend on the parameter $\alpha$ that characterizes the distance between the probability measures.
When the nominal density functions satisfy a symmetry constraint, the two non-linear equations were combined into a single equation. Finally,
the upper bounds on the parameters that control the degree of robustness were derived.
Open problems include proving the monotonicity of the parameters $l_l$ and $l_u$ for increasing $(\epsilon_0,\epsilon_1)$, or $\alpha$.
It was shown that simulation results illustrate the theoretical results.
\appendices
\section{Simplification of $\Phi_1$}\label{appendixa}
From \eqref{eqeq} consider the following steps for
$$\Phi_1=\left(\frac{-1+\lambda_0+\lambda_1+\mu_0+\mu_1-\alpha(-1+\mu_0+\mu_1)}{\lambda_1+\lambda_0 (l/\rho)^{\alpha-1}}\right)^\frac{1}{\alpha-1}$$
\begin{itemize}
\item Dividing the numerator and the denominator by $\lambda_0$ and replacing the term $1+\mu_0/\lambda_0-\alpha\mu_0/\lambda_0$ by $c_1^{\alpha-1}$ results in
$$\Phi_1=\left(\frac{{c_1}^{\alpha-1}+(\lambda_1-1+\mu_1+\alpha- \alpha \mu_1)/\lambda_0}{(\lambda_1/\lambda_0)+ (l/\rho)^{\alpha-1}}\right)^\frac{1}{\alpha-1}.$$
\item Multiplying the numerator and the denominator of the result of the previous step by $\lambda_0/\lambda_1$, replacing the term $1-1/\lambda_1+\mu_1/\lambda_1+\alpha/\lambda_1-\alpha\mu_1/\lambda_1$ by $c_3^{\alpha-1}$ and again multiplying both the numerator and the denominator by $\lambda_1$  gives
$$\Phi_1=\left(\frac{\lambda_0{c_1}^{\alpha-1}+\lambda_1{c_3}^{\alpha-1}}{\lambda_1+\lambda_0(l/\rho)^{\alpha-1}}\right)^\frac{1}{\alpha-1}.$$
\item The result of the previous step is free of parameters $\mu_0$ and $\mu_1$, but still parameterized by $\lambda_0$ and $\lambda_1$. To eliminate them, using the identities $\lambda_0=(1-\alpha)/(c_1^{\alpha-1}-c_2^{\alpha-1})$ and $\lambda_1=(1-\alpha)/(c_4^{\alpha-1}-c_3^{\alpha-1})$ leads to
$$\Phi_1=\left(\frac{{(c_1 c_4)}^{\alpha-1}+{(c_2 c_3)}^{\alpha-1}}{{c_1}^{\alpha-1}-{c_2}^{\alpha-1}+ ({c_4}^{\alpha-1}-{c_3}^{\alpha-1})(l/\rho)^{\alpha-1}}\right)^\frac{1}{\alpha-1}.$$
\item The result from the previous step depends only on $c_1$, $c_2$, $c_3$, $c_4$ and $\alpha$. Using the substitutions $c_1=c_3 l_l$, $c_2=c_4 l_u$ and $c_4=k(l_l,l_u)c_3$ yields
\begin{align}\label{eq37}
&\Phi_1(l,l_l,l_u,c_3;\alpha,\rho)=\nonumber\\
&c_3\left(\frac{k(l_l,l_u)^{\alpha-1}(l_l^{\alpha-1}-l_u^{\alpha-1})}{l_l^{\alpha-1}-(k(l_l,l_u)l_u)^{\alpha-1}+(k(l_l,l_u)^{\alpha-1}-1)(l/\rho)^{\alpha-1}}\right)^\frac{1}{\alpha-1}.
\end{align}
\end{itemize}

\section{Simplification of $\hat\delta$}\label{appendixb}
Since the equivalence of integration domains are given by \eqref{integrationdomains}, only
$$\hspace{-4mm}\hat{\delta}=\frac{\lambda_0(-1+\alpha+\lambda_1+\mu_1-\alpha \mu_1)}{(-1+\alpha)(\lambda_0+\lambda_1 (l/\rho)^{1-\alpha})}-\frac{\lambda_1(\lambda_0+\mu_0-\alpha \mu_0)(l/\rho)^{1-\alpha}}{(-1+\alpha)(\lambda_0+\lambda_1 (l/\rho)^{1-\alpha})},\quad \hat{l}=\rho$$ is required to be simplified. In the following, the simplification is performed in three steps and the domain term $\hat{l}=\rho$ is omitted for the sake of simplicity:
\begin{itemize}
\item Dividing the numerator and the denominator of the first term by $\lambda_1$ and the second term by $\lambda_0$, and replacing the related terms by $c_1^{\alpha-1}$ and $c_3^{\alpha-1}$ results in
\begin{align*}
\hat{\delta}&=\frac{\lambda_0}{-1+\alpha}\cdot\frac{c_3^{\alpha-1}}{\frac{\lambda_0}{\lambda_1}+(l/\rho)^{1-\alpha}}-\frac{\lambda_1}{-1+\alpha}\cdot\frac{c_1^{\alpha-1}(l/\rho)^{1-\alpha}}{1+\frac{\lambda_1}{\lambda_0}(l/\rho)^{1-\alpha}}\\
&=\frac{c_3^{\alpha-1}-c_1^{\alpha-1}(l/\rho)^{1-\alpha}}{(-1+\alpha)\left(\frac{1}{\lambda_1}+\frac{1}{\lambda_0}(l/\rho)^{1-\alpha}\right)}.
\end{align*}
\item The result of the previous step is free of parameters $\mu_0$ and $\mu_1$, but still parameterized by $\lambda_0$ and $\lambda_1$. To eliminate them, using the identities $\lambda_0=(1-\alpha)/(c_1^{\alpha-1}-c_2^{\alpha-1})$ and $\lambda_1=(1-\alpha)/(c_4^{\alpha-1}-c_3^{\alpha-1})$ leads to
$$\hat{\delta}=\frac{(l/\rho)^{1-\alpha}c_1^{\alpha-1}-c_3^{\alpha-1}}{c_4^{\alpha-1}-c_3^{\alpha-1}+(c_1^{\alpha-1}-c_2^{\alpha-1})(l/\rho)^{1-\alpha}}.$$
\item The result from the previous step depends only on $c_1$, $c_2$, $c_3$, $c_4$ and $\alpha$. Using the substitutions $c_1=c_3 l_l$, $c_2=c_4 l_u$ and $c_4=k(l_l,l_u)c_3$ yields
$$\hat{\delta}=\frac{l_l^{\alpha-1}(l/\rho)^{1-\alpha}-1}{(l_l^{\alpha-1}-(k(l_l,l_u)l_u)^{\alpha-1})(l/\rho)^{1-\alpha}+k(l_l,l_u)^{\alpha-1}-1},$$
as wanted.
\end{itemize}

\section*{Acknowledgment}
\label{sec:acknowledge}
The authors would like to sincerely thank the anonymous reviewers for their valuable comments and suggestions to improve the
quality of the paper. This work was supported by the LOEWE Priority Program Cocoon (http://www.cocoon.tu-darmstadt.de).

\bibliographystyle{IEEEtran}
\bibliography{strings4}
\end{document}